\documentclass[10pt,journal]{IEEEtran}
\usepackage{amsmath,amsfonts}
\usepackage{algorithmic}
\usepackage{algorithm}
\usepackage{array}
\usepackage[caption=false,font=footnotesize,labelfont=rm,textfont=rm]{subfig}
\usepackage{textcomp}
\usepackage{stfloats}
\usepackage{url}
\usepackage{verbatim}
\usepackage{graphicx}
\usepackage{cite}
\usepackage{xcolor}
\usepackage{multirow}
\usepackage{adjustbox}
\usepackage{graphicx}
\usepackage{float} 
\usepackage{hyperref}
\usepackage{arydshln}
\usepackage{bbding}
\begin{document}

%Test Time Adaptation for Blind Image Quality Assessment
%StyleAM: Perception-Oriented Unsupervised  Domain Adaption for No-reference Image Quality  Assessment
\title{UPDA: Unsupervised Progressive Domain Adaptation for No-Reference Point Cloud Quality Assessment}

\author{Bingxu~Xie,
    Fang~Zhou,
	Jincan~Wu,
    Yonghui~Liu,
    Weiqing~Li,
	and~Zhiyong~Su% <-this % stops a space
	\thanks{Bingxu Xie, Jincan Wu, Yonghui~Liu, and Zhiyong Su are with the School of Automation, Nanjing University of Science and Technology, Nanjing, Jiangsu Province 210094, P.R. China (e-mail: 2580801219@qq.com, 2500381260@qq.com, 2146623762@qq.com, su@njust.edu.cn)} % <-this % stops a space

     \thanks{Bingxu Xie, and Fang~Zhou are with the National Key Laboratory of Information Systems Engineering, Nanjing, Jiangsu Province 210007, P.R. China (e-mail: 2580801219@qq.com, 406fangzhou@163.com)}
	
	\thanks{Weiqing Li is with the School of Computer Science and Engineering, Nanjing University of Science and Technology, Nanjing, Jiangsu Province 210094, P.R. China (e-mail: li\_weiqing@njust.edu.cn).}
	
	\thanks{Manuscript received 00 00, 0000; revised 00 00, 0000. This work was supported in part by State Key Laboratory of Information Systems Engineering, NO: 05202404. (Corresponding author: Zhiyong Su.)}}

\markboth{Journal of \LaTeX\ Class Files,~Vol.~14, No.~8, August~2021}%
{Shell \MakeLowercase{\textit{et al.}}: A Sample Article Using IEEEtran.cls for IEEE Journals}


\maketitle

\begin{abstract}
While no-reference point cloud quality assessment (NR-PCQA) approaches have achieved significant progress over the past decade, their performance often degrades substantially when a distribution gap exists between the training (source domain) and testing (target domain) data. 
However, to date, limited attention has been paid to transferring NR-PCQA models across domains.
To address this challenge, we propose the first unsupervised progressive domain adaptation (UPDA) framework for NR-PCQA, which introduces a two-stage coarse-to-fine alignment paradigm to address domain shifts. 
At the coarse-grained stage, a discrepancy-aware coarse-grained alignment method is designed to capture relative quality relationships between cross-domain samples through a novel quality-discrepancy-aware hybrid loss, circumventing the challenges of direct absolute feature alignment.
At the fine-grained stage,  a perception fusion fine-grained alignment approach with symmetric feature fusion is developed to identify domain-invariant features, while a conditional discriminator selectively enhances the transfer of quality-relevant features.
Extensive experiments demonstrate that the proposed UPDA effectively enhances the performance of NR-PCQA methods in cross-domain scenarios, validating its practical applicability.
The code is available at \url{https://github.com/yokeno1/UPDA-main}.
\end{abstract}

\begin{IEEEkeywords}
Unsupervised domain adaptation, Point cloud quality assessment, Quality assessment, Progressive alignment.
\end{IEEEkeywords}

\section{Introduction}
\label{sec:intro}

\begin{figure}[!tbp]
\centering
\includegraphics[width = 0.48\textwidth]{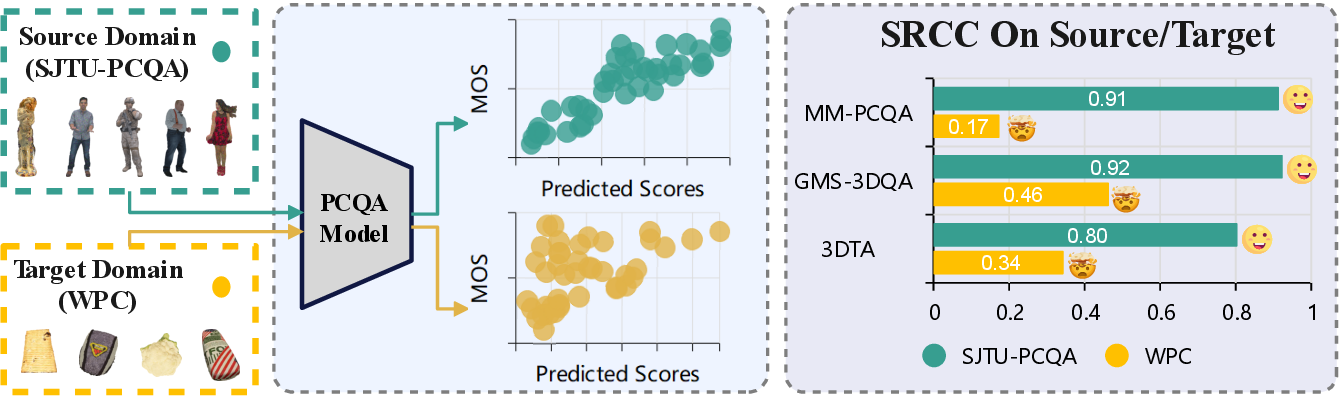}
\caption{Performance degradation problem of typical NR-PCQA methods  (MM-PCQA \cite{zhang2023mm}, GMS-3DQA \cite{zhang2024gms}, and 3DTA \cite{zhu20243dta}) on cross-domain datasets. The source domain (e.g., SJTU-PCQA \cite{yang21sjtu}) and target domain (e.g., WPC \cite{su2019perceptual}) exhibit significant disparities in both content characteristics and distortion types.}
\label{fig:example}
\end{figure}

\IEEEPARstart{W}ith the widespread application of 3D point cloud in augmented/virtual reality (AR/VR) \cite{Akhtar24IFC,Montagud2025VR}, autonomous driving \cite{hao2024rcooper,Khan25LRDNet}, etc., various degradations are inevitable in the process of acquisition, transmission, and processing \cite{Javaheri21Impacts}.
This necessitates the development of effective point cloud quality assessment (PCQA) methodologies to maintain quality of experience (QoE) and optimize performance in point cloud-based systems.
For the past decade, there has been an explosion of PCQA approaches \cite{yang21sjtu,zhang2023mm,zhang2024gms,Shan24gpanet,Su2025PKT,Neri2025Low,zhang2025LargeMultimodal}.
Due to the lack of reference point clouds in real-world applications, no-reference (NR) PCQA methods have gained a lot of attention.
Although learning-based NR-PCQA metrics have achieved excellent performance, they typically require a substantial amount of annotated data for training.
Consequently, they easily suffer from poor generalization ability when there exists a distribution gap between the training and testing domains, which is termed the “domain shift” problem \cite{liu2022survey,fang2024sfdasurvey}.
A typical example is shown in Fig. \ref{fig:example}, where the selected NR-PCQA methods (MM-PCQA \cite{zhang2023mm}, GMS-3DQA \cite{zhang2024gms}, and 3DTA \cite{zhu20243dta}) perform well on the source domain (e.g., SJTU-PCQA \cite{yang21sjtu}), but fail to identify the perceptual quality of the unlabeled target domain (e.g., WPC \cite{su2019perceptual}) due to the domain shift caused by distortion and content differences.
Therefore, a critical yet challenging research question arises: How can we enhance the performance of existing NR-PCQA methods to address the domain shift problem?

Unsupervised domain adaptation (UDA) has attracted appealing attention for tackling the above domain-shift problem through transferring knowledge from a labeled source domain to an unlabeled target domain \cite{kouw2019review,liu2022deep}.
Unfortunately, there is currently no research work conducted on the point cloud distribution gap in the field of PCQA.
To date, only Yang et al. \cite{Yang22itpcqa} have pioneered the UDA technology for PCQA by transferring knowledge from the image domain to the point cloud domain, rather than addressing intra-modal domain shifts between point cloud datasets discussed in Fig. \ref{fig:example}.
On the contrary, UDA has experienced an impressive series of successes in image quality assessment (IQA) \cite{chen21sci,lu2024styleam,li24freqalign,li2024dgqa,chen2021ucda,ou2023troubleshooting,wu2025Underwater}.
The majority of these works \cite{chen21sci,lu2024styleam,li24freqalign,li2024dgqa} adopt a plain one-pass pipeline that globally aligns features only before the quality regression.
% Nevertheless, this one-pass pipeline demonstrates limited alignment capability when processing complex scenes. 
Nevertheless, this one-pass pipeline demonstrates limited alignment capability when processing complex scenes, which are prevalent across various visual tasks.
These complexities can stem from vast differences in scene content (e.g., natural landscapes vs. man-made environments), highly diverse and mixed distortion types, or significant variations in semantic structure.
This limitation has motivated recent research to develop curriculum adaptation frameworks that leverage pseudo-labeling to stratify samples by complexity, enabling sequential alignment for IQA \cite{chen2021ucda,ou2023troubleshooting,wu2025Underwater}.
However, in contrast to 2D images, 3D point clouds exhibit significantly more complex geometric structures and diverse distortion characteristics, presenting unique challenges for quality assessment.
The non-Euclidean, sparse nature of point clouds means that the one-pass global alignment is often too coarse. 
At the same time, the semantic ambiguity and intricate geometries make defining a reliable, universally applicable curriculum for pseudo-labeling exceedingly difficult. 
Therefore, developing UDA techniques for PCQA that explicitly address the inherent structural and distortion characteristics of point clouds represents a crucial research direction.

Inspired by the hierarchical quality assessment mechanism of the human visual system (HVS), which observers first categorize point clouds into coarse quality levels (excellent/average/poor) before assigning fine-grained scores based on local distortion features \cite{Su2025PKT}, we propose the first unsupervised progressive domain adaptation (UPDA) framework for NR-PCQA.
Our approach systematically addresses domain shifts through a biological-motivated, two-stage alignment process: (1) discrepancy-aware coarse-grained alignment (DACA) for relative quality relationship matching, followed by (2) perception fusion fine-grained alignment (PFFA) for absolute quality scoring refinement.
To this end, firstly, the DACA is designed to explore the transferability of pair-wise relative quality relationships by utilizing a coarse-grained ranking task, and adopting the discrepancy-aware maximum mean discrepancy (MMD) to alleviate the discrepancy.
Then, the PFFA is proposed to further explore the transferability of absolute quality scoring capability with a well-designed symmetric feature fusion model and a conditional discriminator.
We summarize our main contributions as follows:
\begin{itemize}
    \item 
    We present the first unsupervised progressive domain adaptation (UPDA) framework for no-reference point cloud quality assessment, designed to address domain shift through a progressive coarse-to-fine alignment strategy. 
    The proposed approach uniquely maintains the original model architecture during inference while only requiring adaptation during training, making it particularly suitable for practical deployment scenarios.
    \item 
    { In the coarse-grained adaptation stage, we introduce a discrepancy-aware coarse-grained alignment (DACA) method that incorporates the concept of discrepancy-awareness into quality ranking loss and MMD loss, aiming to transfer knowledge with reliable ranking signals.
    Specifically, DACA guides the model to prioritize learning the relative relationships between point cloud pairs with large differences in quality, providing the model with more reliable and differentiated cross-domain ranking signals.
    }
    \item 
    {In the fine-grained adaptation stage, we develop a perception fusion fine-grained alignment (PFFA) approach to facilitate cross-domain transfer of absolute quality assessment capabilities. 
    First, we propose a symmetric feature fusion module to enhance the extraction ability of similar features between domains.
    Then, the condition discriminator is improved to adaptively filter quality-sensitive features, which effectively overcomes the limitations of global alignment in existing alignment methods.}

\end{itemize}

\section{Related Works} 
\label{sec:related_works}

\subsection{No-Reference Point Cloud Quality Assessment} 

NR-PCQA aims to estimate point cloud quality without a clean reference point cloud, which can be further divided into traditional and deep learning-based methods.

Traditional NR-PCQA algorithms are mainly based on intrinsic characteristics \cite{chetouani2021deep,zhang2022no,Su2023bit}.
Chetouani et al. \cite{chetouani2021deep} propose to extract three attributes: geometric distance, mean curvature, and gray-level, and then use a convolutional neural network (CNN) to predict quality.
Inspired by the significant success of natural scene statistics (NSS) in IQA tasks, Zhang et al. \cite{zhang2022no} propose a 3D-NSS model that integrates geometric and color attributes and applies NSS to predict quality scores.
Su et al. \cite{Su2023bit} design an evaluation model based on bitstreams for G-PCC without fully decoding compressed PC bitstreams.

Deep learning-based NR-PCQA methods utilize well-designed deep neural networks (DNN) to extract more representative quality-relevant features.
PQA-Net \cite{liu21pqanet} combines multi-view projection and performs distortion classification pre-training to evaluate scores.
GPA-Net \cite{Shan24gpanet} proposes a multi-task graph convolutional network, which considers point cloud content (structure and texture), distortion type, and degree jointly.
GC-PCQA \cite{chen2024gcpcqa} first performs multi-view projection and then constructs a perception-consistent graph to model the mutual dependencies of multi-view projected images.
MM-PCQA \cite{zhang2023mm} and MM-PCQA+ \cite{zhang2025mm+} extract features from both projections and point clouds, and predict quality scores via multi-modal fusion.
% 加入MM-PCQA+文献
Wu et al. \cite{Wu2025Structure} propose a reference-free point cloud quality assessment method based on graph structure and two-stage sampling, which extracts and fuses global and local features separately.
These methods only consider fixed perspectives or limited distortion types, resulting in poor cross-domain evaluation capabilities.
Considering the limitations of existing datasets, Liu et al. \cite{Liu23resscnn} establish a large dataset containing 31 types of distortion and adopt a sparse CNN to extract quality-sensitive features.
Furthermore, to enhance the generalization capability of the model, 3DTA \cite{zhu20243dta} introduces a two-stage sampling method designed to standardize the number of points in the input point cloud and enable efficient quality assessment.
GMS-3DQA \cite{zhang2024gms} proposes a novel multi-projection grid mini-patch sampling strategy to avoid interference from quality-irrelevant information.
D3-PCQA \cite{Liu2025once} decomposes the complex point cloud quality assessment problem into two interpretable and generalizable stages: classification and regression, by introducing a "description domain" and a structured latent space.
{LMM-PCQA \cite{Zhang2024lmmpcqa} explores the feasibility of imparting PCQA knowledge to LMM through text supervision. 
Specifically, this method converts quality labels into text descriptions, enabling LMM to derive quality rating logic from the two-dimensional projection of point clouds.}
% Wang et al. \cite{wang2024zoom} utilize the characteristics of viewpoint changes to extract and learn multi-scale or multi-grained features.

In summary, while significant progress has been made in NR-PCQA, current approaches remain constrained by their fundamental reliance on the assumption of feature space alignment between the target and source domains. 
This strong assumption frequently leads to compromised generalization performance, particularly when evaluating point clouds exhibiting substantial statistical divergence from the training dataset characteristics.

\subsection{Unsupervised Domain Adaptation In Quality Assessment} 

With the help of labeled source data and unlabeled target data, unsupervised domain adaptation (UDA) \cite{Artem19mmd,Eric17adda,sun2016coral,Ganin16dann,Wang23da,chen2021ucda,damodaran2018deepjdot,Wei21Metaalign} aims to eliminate the domain shift between the source domain and the target domain.
Recently, UDA has been widely used in the IQA task to solve the domain shift problem, which can be divided into two categories: one-pass strategy and curriculum adaptation strategy.

The one-pass strategy typically aligns features across domains in a simple and efficient manner, aiming for global alignment.
As an early exploration of the transferability of quality prediction, Chen et al. \cite{chen21sci} utilize MMD to learn the discriminative and transferable pair-wise quality relationship.
To directly explore the transferability of quality prediction tasks, StyleAM \cite{lu2024styleam} finds a more perception-oriented alignment space and designs Style Mixup to increase the consistency between the feature style and its quality score.
FreqAlign \cite{li24freqalign} explores the omni-frequency space to take full advantage of transferable perception knowledge.
To further align the source domain and target domain in complex scenes, several adversarial-based UDA strategies have been combined with curriculum learning.
UCDA \cite{chen2021ucda} divides the target domain into confident and non-confident target subdomains and aligns them through an "easy to hard" curriculum-based learning approach.
Ou et al. \cite{ou2023troubleshooting} propose curriculum domain adaptation, in which the intra-domain quality margin and inter-domain uncertainty are coupled for the curriculum design.

While UDA has become prevalent in image quality assessment, its application to PCQA remains largely unexplored. 
To our knowledge, IT-PCQA \cite{Yang22itpcqa} represents the sole existing work in this direction, which focuses on cross-modal knowledge transfer from images to point clouds by leveraging annotated image quality data. 
Notably, no prior work has specifically addressed cross-domain adaptation between different point cloud domains, which is a critical challenge given the inherent distributional shifts in real-world PCQA scenarios. 
This significant research gap underscores the urgent need to enhance the cross-domain generalization capabilities of NR-PCQA methods.

\begin{figure*}[!htbp]
\centering
\includegraphics[width = \textwidth]{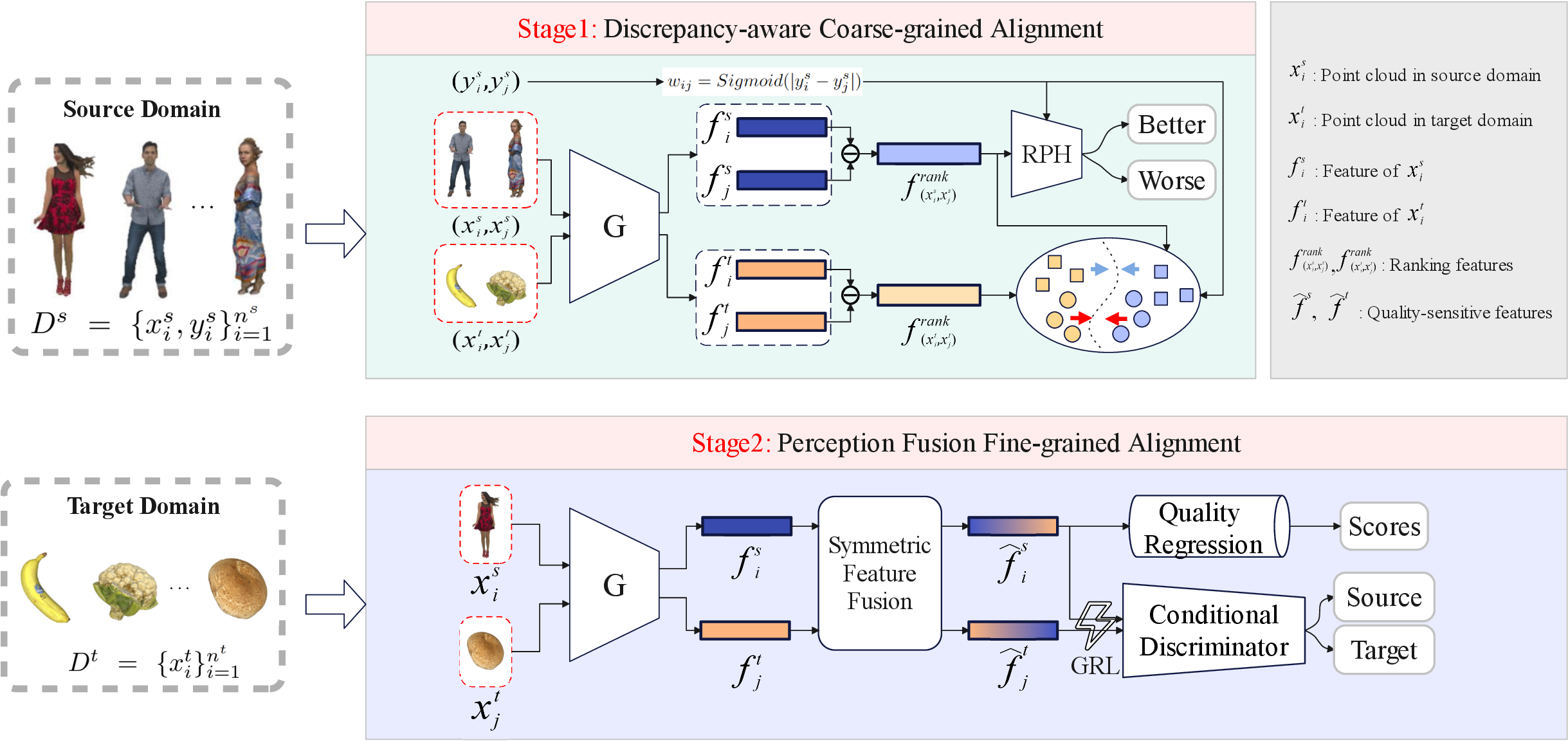}
\caption{{The overall structure of the proposed network. Firstly, in the discrepancy-aware coarse-grained alignment (DACA) stage, we train the feature extractor $G$ by introducing quality difference information. Then, in the perception fusion fine-grained alignment stage,  we implement a hierarchical refinement process to progressively bridge domain gaps while preserving quality-sensitive features by symmetric feature fusion and conditional discriminator.}}
\label{fig:pipeline}
\end{figure*}

\section{Method} 
\label{sec:method}

\subsection{Preliminary}
\label{sec:preliminary}

Unsupervised domain adaptation for PCQA aims to establish cross-domain transferability.
Without loss of generality, we consider conventional PCQA methods to consist of two fundamental components: a feature extractor $G$ and a quality regression head $R$.
Formally, let the labeled source domain dataset be $D^{s}=\{x_{i}^{s}, y_{i}^{s}\}_{i=1}^{n^{s}}$ and the unlabeled target domain dataset be $D^{t}=\{x_{i}^{t}\}_{i=1}^{n^{t}}$.
$x_{i}^{s}$ and $x_{i}^{t}$ denote the sample from the source and target domains, respectively. 
$y_{i}^{s}$ is the ground-truth mean opinion score (MOS) of $x_{i}^{s}$. 
$n^{s}$ and $n^{t}$ are the number of samples in the source and target domains, respectively.
$D^{s}$ and $D^{t}$ have a severe domain gap due to differences in degradation types and content.
To mitigate domain discrepancy, prevailing approaches employ the adversarial-based feature alignment \cite{Ganin15GRL} to optimize domain-invariant representations through minimax optimization, formally expressed as:
\begin{equation}
    \min_{G, R}\max_{D}\mathcal{L}_{Q} - \lambda\mathcal{L}_{D},
\end{equation}
where $\mathcal{L}_{Q}$ denotes the perceptual quality prediction loss on the source domain, $\mathcal{L}_{D}$ represents the domain discriminator loss, and $\lambda$ controls the adaptation intensity. 
Particularly, a discriminator $D$ is optimized to distinguish the source and target domains by applying a domain classification loss. 
% The feature extractor $G$ is enforced to learn the domain-invariant perception knowledge, which is optimized by minimizing the quality aggregation loss $\mathcal{L}_{q}$ with regression head $R$, and maximizing the domain classification loss $\mathcal{L}_{d}$ with the discriminator $D(\cdot)$.
In a word, the feature extractor $G$ is optimized to learn domain-invariant representations through a dual-objective loss function: (1) minimizing the quality regression loss $\mathcal{L}_{Q}$ via the regression head $R$, while (2) maximizing the domain classification loss $\mathcal{L}_{D}$ through the adversarial discriminator $D$. 
This adversarial training strategy forces $G$ to capture quality-relevant features that are robust to domain shifts.

\subsection{Overview}
\label{sec:method-overview}

To tackle the critical challenge of cross-domain distribution shift in PCQA, we propose the first unsupervised progressive domain adaptation (UPDA) framework for NR-PCQA approaches, as illustrated in Fig. \ref{fig:pipeline}.
Our approach operates without requiring MOS annotations in the target domain, thereby overcoming a key limitation in real-world PCQA deployment.
The UPDA framework introduces a novel two-stage progressive alignment strategy comprising discrepancy-aware coarse-grained alignment (DACA) and perception fusion fine-grained alignment (PFFA).
The proposed two-stage progressive alignment strategy fundamentally differs from conventional UDA schemes for IQA that directly minimize discrepancies in absolute quality feature distributions \cite{wu2025Underwater} - a methodology particularly susceptible to failure when source and target domains exhibit substantial distributional divergence.

{In the coarse-grained adaptation stage, DACA introduces quality difference information into ranking loss and MMD loss to optimize the alignment process.}
With the DACA, we can train a coarse-grained feature extractor $G$ to mitigate quality ranking bias in the target domain.
In the fine-grained adaptation stage, building upon the coarse alignment, we propose the PFFA to enhance precise score prediction ability.
{PFFA focuses on aligning quality-related features, effectively eliminating interference caused by introducing quality-irrelevant features in global alignment.}
A symmetric feature fusion module is first introduced to identify domain-invariant quality attributes for alignment.
Subsequently, a quality regression and a conditional discriminator are designed to transfer the absolute quality prediction capability to the target domain.

\subsection{Discrepancy-aware Coarse-grained Alignment}

{The DACA aims to utilize the quality difference information between point cloud pairs to achieve coarse-grained alignment.
% The DACA aims to exploit the relative quality ranking relationship between point cloud pairs to achieve coarse-grained alignment.
We argue that the adaptation process should prioritize learning and aligning point cloud pairs with significant quality discrepancies and a certain ranking relationship, as these point cloud pairs provide the clearest and most robust cross-domain quality relationship signals.
Therefore, we introduce a weighting mechanism driven by quality discrepancies to guide the model to focus on point cloud pairs with significant quality discrepancies.}
% {Specifically, DACA introduces the difference measure into the ranking loss and MMD loss, which is different from the traditional ranking-based adaptation method.}
% Unlike conventional approaches that treat all sample pairs equally, we argue that pairs with significant quality discrepancies provide more reliable ranking signals and should be prioritized during both ranking feature learning and alignment.
% Therefore, introducing the weight $w_{ij}$ into the loss allows the model to focus on learning pairs with significant quality discrepancies, thereby enhancing the discriminative ability and alignment effectiveness of ranking features.
Specifically, we design a weighting factor $w_{ij}$ to perform weighting in the process of alignment.
The DACA first learns discriminative ranking features from the source domain through a dedicated discrepancy-aware ranking prediction module.
Then, leveraging the learned ranking features, we design a discrepancy-aware maximum mean discrepancy module to transfer these ranking-aware representations to the target domain.
% The traditional ranking loss treats all sample pairs equally. 
% We recognize that samples with significant differences in quality can provide clearer and more reliable ranking signals. 
% Therefore, introducing the weight $w_{ij} $ into the ranking loss allows the model to focus on learning pairs with significant quality discrepancy, thereby enhancing the discriminative power of the ranking features.

\subsubsection{{Discrepancy-aware Ranking Prediction}}

The discrepancy-aware ranking prediction module is adopted to learn ranking feature $f_{(x_{i}^{s}, x_{j}^{s})}^{rank}$ which represents the quality relationship between point cloud pairs $(x_{i}^{s}, x_{j}^{s})$ in the source domain $D^{s} = \{x_{i}^{s}, y_{i}^{s}\}_{i=1}^{n^{s}}$.
{Considering that binary labels (better/worse) only retain the "direction" of quality order and completely ignore the "magnitude" of quality discrepancies, we argue that not all ranking signals are equally important.}

First, let the output of the feature extractor $G$ be $f = G(x)$, the ranking feature $f_{(x_{i}^{s}, x_{j}^{s})}^{rank}$ can be learned by measuring the differences between the corresponding features $f_{i}^{s}$ and $f_{j}^{s}$ as follows:
\begin{equation}
    f_{(x_{i}^{s}, x_{j}^{s})}^{rank} = f_{i}^{s} - f_{j}^{s}.
\end{equation}
Then, the probability of the ranking classification $P_{(x_{i}^{s},x_{j}^{s})}^{rank}$ can be estimated as follows:
\begin{equation}
    \begin{aligned}{P_{(x_{i}^{s},x_{j}^{s})}^{rank}}&=P\left((l_{ij}^{s}=1)|(x_{i}^{s},x_{j}^{s})\right)\\&=P\left((y_{i}^{s}>y_{j}^{s})|(x_{i}^{s},x_{j}^{s})\right), \end{aligned}
\end{equation}
where $P(\cdot)$ represents the probability, and $l_{ij}^{s}$ denotes the ranking label between $y_{i}^{s}$ and $y_{j}^{s}$, indicating the relative quality order of $x_{i}^{s}$ and $x_{j}^{s}$.
Specifically, $l_{ij}^{s} = 1$ if $y_{i}^{s} > y_{j}^{s}$ ($x_{i}^{s}$ has higher quality than $x_{j}^{s}$), and $l_{ij}^{s} = 0$ otherwise.
The ranking prediction head $RPH$ which consists of fully connected ($FC$) layers and a $Softmax$ layer are used to compute $P_{(x_{i}^{s},x_{j}^{s})}^{rank}$ as follows:
\begin{equation}
P_{(x_{i}^{s},x_{j}^{s})}^{rank} = RPH(f_{(x_{i}^{s},x_{j}^{s})}^{rank}) = Softmax(FC(f_{(x_{i}^{s},x_{j}^{s})}^{rank})).
\end{equation}
Finally, the discrepancy-aware binary cross-entropy loss is employed to train the feature extractor $G$ to preserve relative quality ordering between point cloud samples:
\begin{equation}
\begin{aligned}
\mathcal{L}_{\mathrm{D-rank}} = -&\frac{1}{\sum_{i=1}^{n_s}\sum_{j=1}^{n_s} w_{ij}}\sum_{i=1}^{n^{s}}\sum_{j=1}^{n^{s}} w_{ij} \cdot [l_{ij}^{s}\cdot\operatorname{log}P_{(x_{i}^{s},x_{j}^{s})}^{rank} \\ &+ (1-l_{ij}^{s})\cdot\operatorname{log} (1-P_{(x_{i}^{s},x_{j}^{s})}^{rank})],
\end{aligned}
\end{equation}
where weight $w_{ij}$ is defined as:
\begin{equation}
\label{eq:w}
w_{ij} = Sigmoid(|y_i^s - y_j^s|).
\end{equation}

\subsubsection{{Discrepancy-aware Feature Adaptation}}

{The discrepancy-aware feature adaptation module is designed to facilitate the transfer of knowledge with reliable ranking relationships to the target domain. 
It proposes discrepancy-aware maximum mean discrepancy (D-MMD) as a distribution matching constraint to minimize the cross-domain discrepancy between source and target feature distributions.}
The traditional MMD has inherent flaws when applied to sorting tasks: it assumes that the weights of all sample pairs are equal, without considering the relative differences in quality scores.  
The MMD is incapable of discriminating between sample pairs exhibiting substantial and slight quality differences.
Consequently, it fails to preserve the gradient information essential for accurate ranking.
The proposed D-MMD minimizes the cross-domain discrepancy by incorporating quality differences as weighting factors, which effectively encourages the model to learn significant discriminative information.

Given a pair of point clouds $(x_{i}^{t}, x_{j}^{t})$ from the target domain $D^{t}$, their ranking feature $f_{(x_{i}^{t}, x_{j}^{t})}^{rank}$ can be defined as:
\begin{equation}
    f_{(x_{i}^{t}, x_{j}^{t})}^{rank} = f_{i}^{t} - f_{j}^{t}.
\end{equation}
Subsequently, the proposed D-MMD is employed as a loss to reduce the discrepancy between the source and target domains:
\begin{align}
\nonumber
\mathcal{L}_{D-MMD} = &\| \frac{\sum_{i=1}^{n^s}\sum_{j=1}^{n^s} w_{ij} \cdot \phi( f_{(x_{i}^{t}, x_{j}^{t})}^{r a n k} )}{\sum_{i=1}^{n_s}\sum_{j=1}^{n_s} w_{ij}} \\ &- \frac{1} {n^{t}} \sum_{i=1}^{n^t}\sum_{j=1}^{n^t} \phi( f_{(x_{i}^{t}, x_{j}^{t})}^{r a n k} ) \|_{\mathcal{H}}^{2}, 
\end{align}
where $\phi (\cdot)$ is a function that maps the features into the Reproducing Kernel Hilbert Space (RKHS) \cite{Arthur12rkhs}.

Finally, in the DACA stage, we can train a coarse-grained feature extractor $G$ to mitigate quality ranking bias in the target domain by minimizing the loss function as follows:
\begin{equation}
\begin{gathered}
        {\cal{L}}_{DACA} = {\cal{L}}_{D-rank} + \nu {\cal{L}}_{D-MMD},
\end{gathered}
\end{equation}
where $\nu$ is the weighting factor.
The feature extractor $G$ produces representations encoding preliminary coarse-grained quality rankings, which subsequently undergo refinement in the PFFA stage.

\subsection{Perception Fusion Fine-grained Alignment}

Building upon the DACA-pretrained feature extractor $G$ that learns coarse-grained relative quality representations, we propose the PFFA to further enhance $G$'s capability for absolute quality prediction. 
Unlike conventional IQA adaptation approaches \cite{lu2024styleam,li24freqalign,ou2023troubleshooting} that directly feed $G$'s features to both domain discriminator $D$ and regression head $R$, resulting in insufficient alignment caused by global adaptation, the PFFA implements a hierarchical refinement process to progressively bridge domain gaps while preserving quality-sensitive features.
Generally speaking, the PFFA implements the following process: (1) A symmetric feature fusion module first identifies and fuses quality-relevant features across domains through cross-attention mechanisms, followed by (2) a regression head for predicting quality scores and a conditional discriminator with adaptive thresholding that selectively aligns domain-invariant features most beneficial for quality prediction. 
Throughout this process, the feature extractor $G$ and regression head $R$ are jointly optimized using labeled source data to maintain robust absolute quality assessment performance.

\subsubsection{Symmetric Feature Fusion}

The symmetric feature fusion employs a symmetric cross-attention transformer block to dynamically identify and enhance shared feature representations between source feature $f^{s}_{i}$ and target feature $f^{t}_{j}$, where both $f^{s}_{i}$ and $f^{t}_{j}$ are generated by the feature extractor $G$.
The multi-head cross-attention ($MCA$) \cite{vaswani2017attention} is employed as the fundamental structure of the symmetric cross-attention transformer block.
Explicitly, given the three common sets of inputs: query set $Q$, key set $K$, and value set $V$, we define the $MCA$ operation as:
\begin{equation}
    \begin{gathered}
        Q_i = QW_{i}^{Q},\\
        K_i = KW_{i}^{K},\\
        V_i = VW_{i}^{V}, \\
        h_{i} = Softmax(Q_{i}K_{i}^T/\sqrt{d})V_{i}, \\
        MCA(Q,K,V) = Concact( h_{1}, h_{2}..., h_{n})W^H,
    \end{gathered}
\end{equation}
where $h_{i}$ is the $i$-th head, and $W^H, W^Q, W^K, W^V$ are learnable linear mappings.
The final quality-sensitive features can be represented as follows:
\begin{equation}
\begin{gathered}
    \hat{f}^{s}_{i} = f^{s}_{i} + MCA(f^{s}_{i},f^{t}_{j},f^{t}_{j}),\\
    \hat{f}^{t}_{j} = f^{t}_{j} + MCA(f^{t}_{j},f^{s}_{i},f^{s}_{i}).
\end{gathered}
\end{equation}

\subsubsection{Conditional Discriminator}

The conditional discriminator selectively retains fusion features that contribute to quality regression.
To minimize the domain discrepancy across source and target domains, the domain discriminator is trained with the help of a gradient reverse layer (GRL) \cite{Ganin15GRL}.
However, commonly used domain discriminators ignore the side effects of partial negative features (i.e., the partial features that have bad correlations with their quality scores) \cite{lu2024styleam}.
Inspired by \cite{Yang22itpcqa}, we relax the alignment by revising the discriminator loss as:
\begin{equation}
    \begin{gathered}
    \mathcal{L}_{D} = \mathcal{L}_{adv}^{s}+\mathcal{L}_{adv}^{t}, \\
    \mathcal{L}_{adv}^{s}=-\frac{1}{n^{s}}\sum_{i=1}^{n^{s}}\log(|D(\hat{f}_{i}^{s})-h|), \\
    \mathcal{L}_{adv}^{t}=-\frac{1}{n^{t}}\sum_{j=1}^{n^{t}}\log(1-D(\hat f_{j}^{t})),
    \end{gathered}
\end{equation}
where $h$ denotes whether the alignment needs to be relaxed. 
We use the mean squared error (MSE) to measure whether the fused features $\hat{F}^{s}$ contribute to quality regression.
The definition of $h$ can be written as:
\begin{equation}
        h=
    \begin{cases}
    0,\quad \mathrm{if} \ \  MSE(R(\hat{f}^{s}_{i}),y^{s}_{i}) > MSE(R({f}^{s}_{i}),y^{s}_{i}) \\
    1,\quad \mathrm{others} & 
    \end{cases}
\end{equation}
where $R(\cdot)$ is the quality regression head, and $y^{s}_{i}$ is the ground-truth score.

\subsubsection{Quality Regression}

The quality regression is adopted to regress the quality representation into predicted perceptual quality scores.
Following common practice, we use MSE as the loss function to keep the predicted score close to the ground-truth MOS, which can be derived as:
\begin{equation}
    \mathcal{L}_{Q} = \frac{1} {n^s} \sum_{i=1}^{n^s} ( q_{i}-y_{i}^{s} )^{2}, 
\end{equation}
where $q_{i}$ is the predicted quality score, $y_{i}^{s}$ is the ground-truth MOS.

Finally, in the PFFA stage, we can train the feature extractor $G$ and regression head $R$ of given PCQA methods through the following loss:
\begin{equation}
\begin{gathered}
        {\cal{L}}_{PFFA} = {\cal{L}}_{Q} + \mu \mathcal{L}_{D},
\end{gathered} 
\end{equation}
where $\mu$ is the weighting factor.
Note that the feature extractor $G$ is pretrained at the DACA stage.
After that, the well-trained PCQA methods can be used to predict absolute quality scores of point clouds in the target domain.

\section{Experiments} 

\subsection{Datasets and Criteria}

Three popular PCAQ datasets are utilized to validate the performance of the proposed UPDA framework, which include SJTU-PCQA \cite{yang21sjtu}, WPC \cite{su2019perceptual}, and WPC 2.0 \cite{Liu21wpc2}. 
\begin{itemize}
    \item The SJTU-PCQA dataset \cite{yang21sjtu} contains 9 reference point clouds. Each reference point cloud is subject to 7 types of distortions at 6 different intensities, including octree-based compression, color noise, geometry gaussian noise, downsampling, and their three combinations. Therefore, the dataset generates a total of 378 distorted point clouds.
    \item The WPC dataset \cite{su2019perceptual} contains 20 reference point clouds. Four types of distortion are introduced for each reference point cloud, including downsampling, Gaussian noise, MPEG-GPCC compression, and MPEG-VPCC compression, resulting in 740 distorted point clouds.
    \item The WPC 2.0 dataset \cite{Liu21wpc2} contains 16 reference point clouds, which are downgraded using 25 different V-PCC (Video Point Cloud Compression) settings, resulting in 400 distorted point clouds.
\end{itemize}

Two mainstream evaluation criteria in the quality assessment field are utilized to compare the correlation between the predicted scores and ground-truth MOS, which include Spearman Rank Correlation Coefficient (SRCC) \cite{spearman1961srcc} and Pearson Linear Correlation Coefficient (PLCC) \cite{pearson1895plcc}. 
%A top-performing model should have SRCC and PLCC values that approach 1.
To address the scale discrepancies between predicted quality scores and their corresponding subjective ratings, a five-parameter logistic function \cite{antkowiak2000final} is employed to align the predicted scores with the subjective MOS:
\begin{equation}
    \hat{y}=\beta_{1} \left( 0. 5-\frac{1} {1+e^{\beta_{2} ( y-\beta_{3} )}} \right)+\beta_{4} y+\beta_{5}, 
\end{equation}
where $\left\{\beta_{i}| i = 1, 2, \cdots, 5 \right\}$ are parameters to be fitted, $y$ and $\hat{y}$ are the predicted scores and mapped scores, respectively.

%No-reference Screen Content Image Quality Assessment with Unsupervised Domain Adaptation
\subsection{Implementation Details}

\begin{table*}[!tbp]
\caption{Performance comparison with SOTA PCQA metrics on the cross-distortion scenarios. Type $1-7$ represents octree-based compression, color noise, downsampling, downsampling \& color noise, downsampling \& geometry gaussian noise, geometry gaussian noise, and color noise \& geometry gaussian noise, respectively.}
\label{cross-distortion}
\resizebox{\linewidth}{!}{
\begin{tabular}{lccccccccccccccc}
\hline
\multirow{2}{*}{Backbone} & \multirow{2}{*}{Method}  & \multicolumn{2}{c}{Type 1} & \multicolumn{2}{c}{Type 2} & \multicolumn{2}{c}{Type 3} & \multicolumn{2}{c}{Type 4} & \multicolumn{2}{c}{Type 5} & \multicolumn{2}{c}{Type 6} & \multicolumn{2}{c}{Type 7} \\
   &          & SRCC         & PLCC        & SRCC         & PLCC        & SRCC         & PLCC        & SRCC         & PLCC        & SRCC         & PLCC        & SRCC         & PLCC        & SRCC         & PLCC        \\ \hline
PQA-Net \cite{liu21pqanet}  & NoAdapt    & 0.1573       & 0.1762      & 0.5782       & 0.6213      & 0.3745       & 0.4012      & 0.5623       & 0.5213      & 0.3771       & 0.4231      & 0.4854       & 0.4908      & 0.6792       & 0.7012      \\
IT-PCQA \cite{Yang22itpcqa}  & UDA     & 0.7321       & 0.7628      & 0.6732       & 0.6281      & 0.6893       & 0.7001      & 0.8856       & 0.8923      & 0.8542       & 0.8467      & 0.8152       & 0.7982      & 0.7872       & 0.7763      \\ \hdashline

\multirow{3}{*}{GMS-3DQA \cite{zhang2024gms}}  & NoAdapt       & 0.8224       & 0.7992      & 0.7887       & 0.7551      & 0.6421       & 0.6886      & 0.9256       & 0.9005      & 0.9166       & 0.9372      & 0.8803       & 0.9497      & 0.9438       & 0.9551      \\
& DirAdapt     & 0.8127       & 0.7881      & 0.7903       & 0.7673      & 0.6519       & 0.6782      & 0.9356       &\textbf{ 0.9583}      & 0.9201       & 0.9416      & 0.9202      & 0.9467      & 0.9532       & \textbf{0.9656}      \\
 & UPDA    & \textbf{0.8601}       & \textbf{0.8211}      & \textbf{0.8446 }      & \textbf{0.8353}      & \textbf{0.7452}       & \textbf{0.8419 }     & \textbf{0.9499}       & 0.9493      & \textbf{0.9476 }      & \textbf{0.9683}      & \textbf{0.9263 }      & \textbf{0.9518}      & \textbf{0.9632 }      & 0.9598      \\ \hdashline
\multirow{3}{*}{3DTA \cite{zhu20243dta}}  &  NoAdapt         & 0.8108       & 0.8650      & 0.7855       & 0.7824      & 0.9097       & 0.9458      & \textbf{0.9571}       & 0.9568      & 0.9536       & 0.9410      & 0.8845       & 0.8576      & 0.9142       & 0.9246      \\
& DirAdapt      & 0.7983       & 0.8247      & 0.7706       & 0.7857      & 0.9105       & \textbf{0.9564}      & 0.9548       & \textbf{0.9663}     & 0.9546       & 0.9483      & 0.8902      & 0.8674      & 0.9205       & 0.9435     \\
  &  UPDA   & \textbf{0.8316}       & \textbf{0.8859 }     & \textbf{0.8518}       & \textbf{0.8475}      & \textbf{0.9207}       & 0.9498      & 0.9531       & 0.9623      & \textbf{0.9621}       & \textbf{0.9727}      & \textbf{0.9376}       & \textbf{0.9495}      & \textbf{0.9587}       & \textbf{0.9639}      \\ \hdashline
  \multirow{3}{*}{MM-PCQA \cite{zhang2023mm}}    &  NoAdapt       & 0.7595       & 0.8018      & 0.6413       & 0.6571      & \textbf{0.7266}       & \textbf{0.8401}      & 0.9365       & 0.9422      & 0.9108       & 0.9175      & 0.8844       & 0.9145      & 0.8953       & 0.9337      \\
& DirAdapt   & 0.7483       & 0.7658      & 0.6032      & 0.6145      & 0.7012      & 0.8019      & 0.9363       & \textbf{0.9532}      & 0.9276       & 0.9201     & \textbf{0.8994}       & 0.9016      & 0.9034       & 0.9415       \\
  & UPDA    & \textbf{0.7708}       & \textbf{0.8125 }     & \textbf{0.6492}       & \textbf{0.6712}      & 0.5930       & 0.7182      & \textbf{0.9423}       & 0.9501      & \textbf{0.9365}       & \textbf{0.9479}      & 0.8718       & \textbf{0.9315}      & \textbf{0.9364}       & \textbf{0.9572 }     \\ \hline
\end{tabular}}

\end{table*}

The proposed framework is implemented in PyTorch, and all the experiments are conducted on a 24 GB NVIDIA RTX 4090 GPU.
We apply our UPDA to three state-of-the-art (SOTA) NR-PCQA methods, namely 3DTA \cite{zhu20243dta}, GMS-3DQA \cite{zhang2024gms}, and MM-PCQA \cite{zhang2023mm}. 

%No-reference Screen Content Image Quality Assessment with Unsupervised Domain Adaptation
Throughout all experiments, we empirically set $\nu = 1$ for ${\cal{L}}_{DACA}$ and $\mu = 0.8$ for ${\cal{L}}_{PFFA}$.
% To train the feature extraction $G$ and regression head $R$, we follow the settings in the original paper.
To prevent memory overflow, we have uniformly reduced the batch size to half of the original method's settings, as the network needs to process data from both the source and target domains simultaneously.
In DACA, we follow the original settings of each method and adjust the necessary parameters for coarse-grained training with a total epoch of 50.
Specifically, for 3DTA \cite{zhu20243dta}, we adopt a stochastic gradient descent (SGD) optimizer with a learning rate of $1e-4$ and a batch size of 16 for training.
For GMS-3DQA \cite{zhang2024gms}, the Adam optimizer is employed with the $1e-4$ initial learning rate, and the default batch size is set as 16.
For MM-PCQA \cite{zhang2023mm}, the Adam optimizer is utilized with an initial learning rate of $5e-5$, and the batch size is set as 4.
During PFFA, we follow the optimizer in DACA and reduce the learning rate to $1e-5$ with a total epoch number of 50.

For all datasets, we linearly rescale the quality scores to a common range $[0,10]$.
We first divide the target dataset into training and testing sets in a $8:2$ ratio. 
Subsequently, we align the training set of target domain with the complete labeled dataset of source domain, and the performance is evaluated on the testing set of target domain. 
Given that the SJTU-PCQA, WPC, and WPC 2.0 datasets contain 9, 20, and 16 independent groups of point clouds, respectively, we perform 9-fold, 5-fold, and 4-fold cross-validation on the three datasets accordingly.
The average performance is recorded as the final results.
It is worth mentioning that there is no content overlap between the training and testing sets.

\subsection{Performance Comparison}

We select three SOTA models (3DTA \cite{zhu20243dta}, GMS-3DQA \cite{zhang2024gms}, and MM-PCQA \cite{zhang2023mm}) as backbones, and evaluate the proposed UPDA with two variants (NoAdapt, DirAdapt).
The NoAdapt indicates that the model is trained on the source domain and directly tested on the target domain without adaptation.
The DirAdapt is a direct application of the adversarial learning \cite{Ganin16dann}, which means that a domain discriminator is added after the feature extractor $G$ of the model for adaptation.
The UPDA first adds the DACA for coarse-grained adaptation after the $G$, and then the PFFA replaces the DACA for fine-grained adaptation.
Besides, two PCQA methods (including a learning-based NR-PCQA method PQA-Net \cite{liu21pqanet}, and a UDA method IT-PCQA \cite{Yang22itpcqa}) are also selected for performance comparison.
For fair comparison, we adjust the source domain of IT-PCQA \cite{Yang22itpcqa} from the image domain to the point cloud domain.

\subsubsection{Performance Under the Cross-distortion Scenario}

We conduct comprehensive experiments under cross-distortion scenarios characterized by significant distortion discrepancies, where the source and target domains exhibit substantial variations in distortion profiles (e.g., compression algorithms, noise patterns). 
Specifically, we use the SJTU-PCQA dataset \cite{yang21sjtu} and divide the point clouds into seven distinct groups based on the types of distortions present in the dataset. 
Each group represents a specific distortion category, such as noise, compression, or downsampling, which are common challenges in real-world applications.
To systematically evaluate cross-distortion generalization, we employ a leave-one-distortion-type-out protocol. 
In this framework, we iteratively: (1) designate point clouds containing a single distortion type as the target domain, while (2) utilizing all remaining samples as the source domain. 
This rigorous evaluation paradigm enables quantitative assessment of our method's ability to generalize to completely novel distortion types. 
By exhaustively testing all possible distortion-type exclusions, we obtain a comprehensive generalization profile across diverse distortion scenarios. 
Such thorough validation is particularly critical for real-world deployment, where point clouds frequently exhibit complex, mixed distortion patterns that may not be represented in training data.

The experimental results from Table \ref{cross-distortion} indicate that the backbones exhibit significant performance differences across different types of distortions, reflecting their sensitivity to specific distortion types. 
In contrast, three models (3DTA \cite{zhu20243dta}, GMS-3DQA \cite{zhang2024gms}, and MM-PCQA \cite{zhang2023mm}) using the UPDA strategy exhibit stable cross-distortion generalization ability: achieving an average performance improvement of 3.2\%/2.8\% for SRCC/PLCC in seven types of distortion tasks.
In particular,  GMS-3DQA \cite{zhang2024gms} with UPDA improves SRCC from 0.6421 to 0.7452 and PLCC from 0.6886 to 0.8419 on the Type 3 distortion (downsampling), verifying the effectiveness of the progressive domain alignment strategy.
In mixed-distortion scenarios where source and target domains share similar distortion compositions, the inherent domain discrepancy is naturally reduced, theoretically limiting potential cross-domain performance gains. 
Nevertheless, our experimental results demonstrate that UPDA's progressive feature alignment mechanism still achieves measurable improvements. 
For example, when applied to GMS-3DQA \cite{zhang2024gms}, UPDA elevates SRCC performance from 0.9256 to 0.9499 (an absolute improvement of 2.71\%) on Type 4 distortions (downsampling \& color noise). 
This evidence suggests that UPDA can effectively extract transferable cross-domain features and enhance model generalization, even in scenarios with minimal domain shift.

\subsubsection{Performance Under the Cross-dataset Scenario}

\begin{table*}[!ht]
\caption{Performance comparison with SOTA PCQA metrics under the cross-dataset scenario. Note that the UDA method IT-PCQA \cite{Yang22itpcqa} performs cross-modal adaptation by transferring knowledge from the image domain to the point cloud domain, rather than addressing intra-modal domain shifts between point cloud datasets.}
\label{cross-dataset}
\centering
\resizebox{\linewidth}{!}{
\begin{tabular}{lccccccccccccc}
\hline
       \multirow{2}{*}{Backbone}                         & Train/Test & \multicolumn{2}{c}{SJTU/WPC} & \multicolumn{2}{c}{SJTU/WPC 2.0} & \multicolumn{2}{c}{WPC/SJTU} & \multicolumn{2}{c}{WPC/WPC 2.0} & \multicolumn{2}{c}{WPC 2.0/SJTU} & \multicolumn{2}{c}{WPC 2.0/WPC} \\ 
             &   Method              & SRCC       & PLCC       & SRCC         & PLCC        & SRCC          & PLCC          & SRCC         & PLCC        & SRCC          & PLCC          & SRCC       & PLCC       \\ \hline
PQA-Net \cite{liu21pqanet}      &       NoAdapt                  & 0.3518     & 0.3588     & 0.3684       & 0.3770      & 0.6267        & 0.6610        & 0.3528       & 0.3313      & 0.6374        & 0.6323        & 0.3302     & 0.3145     \\
% ResSCNN$^{\ast}$         &    NoAdapt         & 0.2580     & 0.2690     & 0.4800       & 0.4700      & 0.5350        & 0.5720        & -            & -           & 0.6000        & 0.6700        & -          & -          \\
% GPA-Net$^{\ast}$       &   NoAdapt         & 0.4240     & 0.4310     & -            & -           & 0.5350        & 0.5740        & -            & -           & -             & -             & -          & -          \\
IT-PCQA  \cite{Yang22itpcqa}    &     UDA        & 0.4219     & 0.4233     & 0.3892       & 0.3910      & 0.4923        & 0.4875        & 0.7993       & 0.8102      & 0.5423        & 0.5376        & 0.5018     & 0.5129     \\ \hdashline
\multirow{3}{*}{GMS-3DQA \cite{zhang2024gms}} & NoAdapt        & 0.4602     & 0.4951     & 0.2883       & 0.3879      & 0.7592        & 0.7942        & 0.8025       & 0.7952      & 0.7154        & 0.7822        & 0.6139     & 0.6117     \\
& DirAdapt              & 0.4435   & 0.4327     & 0.3012       & 0.3687     & 0.7621       & 0.7592       & 0.8112       & 0.8074      & 0.6954        & 0.7362       & 0.6223    & 0.6018   \\ 
                          & UPDA             & \textbf{0.5397}     & \textbf{0.5542 }    & \textbf{0.5124}      & \textbf{0.5207}      & \textbf{0.8167}        & \textbf{0.8354}        & \textbf{0.8619}       & \textbf{0.8643}      & \textbf{0.7398 }       & \textbf{0.7753 }       & \textbf{0.6613}     & \textbf{0.6674}     \\ \hdashline
\multirow{3}{*}{3DTA \cite{zhu20243dta}}     & NoAdapt        & 0.3372     & 0.4409     & 0.4549       & 0.5038      & 0.7630        & 0.8046        & 0.8711       & 0.8682      & 0.7298        & 0.7184        & \textbf{0.5929}     & 0.5967     \\
& DirAdapt             & 0.3564     & 0.4543     & 0.4783       & 0.4956     & \textbf{0.7839 }      & 0.8114        & \textbf{0.8972}      & \textbf{0.8806}     & 0.7125        & 0.7002       & 0.5913     & 0.6019   \\ 
                          & UPDA             &\textbf{ 0.4814 }    & \textbf{0.5483}     & \textbf{0.5774}       & \textbf{0.6238}      & 0.7819        & \textbf{0.8189}        & 0.8792       & 0.8641      & \textbf{0.7523}        & \textbf{0.7401}        & 0.5914     & \textbf{0.6352 }    \\ \hdashline
\multirow{3}{*}{MM-PCQA \cite{zhang2023mm}}  & NoAdapt         & 0.1752     & 0.4580     & 0.3127       & 0.3363      & 0.7851        & 0.8238        & 0.8480       & 0.8544      & 0.6587        & 0.7517        & 0.5312     & 0.5523     \\
& DirAdapt             & 0.1852     & 0.4231     & 0.3263      & 0.3018      & 0.7856        & 0.8094        & \textbf{0.8602 }      & 0.8612      & 0.6544        & 0.7610       & 0.5117    & 0.5290    \\ 
 & UPDA             & \textbf{0.2381}     & \textbf{0.4659}     & \textbf{0.3812}       &\textbf{ 0.3793}      & \textbf{0.8173}        & \textbf{0.8283 }       & 0.8546       & \textbf{0.8620 }     & \textbf{0.6937}        & \textbf{0.7842}        & \textbf{0.5619}     & \textbf{0.5791 }    \\ \hline
\end{tabular}}
\end{table*}

In addition to cross-distortion scenarios, we further explore cross-dataset scenarios to evaluate the effectiveness of our method under more challenging situations.
We conduct comprehensive experiments under cross-dataset scenarios characterized by significant domain discrepancies, where the source and target domains exhibit substantial variations in content characteristics (e.g., object categories, geometric complexity) and distortion profiles. 
As evidenced in Table \ref{cross-dataset}, we draw the following conclusions.

% Firstly, we can observe that the backbones with variant NoAdapt exhibit severe performance degradation when they are directly applied across datasets.
Firstly, we can observe that the backbones with variant NoAdapt exhibit poor performance when they are directly applied across datasets.
While GMS-3DQA \cite{zhang2024gms} addresses content diversity through grid-based mini-patch sampling for quality map generation, and 3DTA \cite{zhu20243dta} employs a two-stage adaptive sampling strategy for variable point cloud inputs, neither approach considers the critical challenge of domain shift in cross-dataset scenarios. 
This fundamental limitation inevitably leads to significant performance degradation when deployed across heterogeneous datasets.
It is particularly noteworthy that the performance collapse (SRCC = 0.2883, PLCC = 0.3879) is observed when transferring GMS-3DQA \cite{zhang2024gms} from SJTU-PCQA to WPC 2.0, underscoring the limitations of fixed projection strategies in handling domain-specific distortion characteristics. 

Secondly, our experiments reveal that three DirAdapt-variant backbones underperform the NoAdapt baseline in most test scenarios. 
{The possible reason is that domain differences caused by point cloud content and distortion types can lead to insufficient alignment when DirAdapt forces global adaptation.
For example, the distortion types in the SJTU-PCQA dataset do not include MPEG-GPCC and MPEG-VPCC in the WPC dataset, which introduces a domain difference gap.}
{The sole exception occurs in the WPC/WPC 2.0 configuration, where significant overlap exists between the domain distributions of content and distortion.}
% {This is because there is content overlap and the same distortion between domains.}
% This finding suggests that straightforward feature alignment can degrade model discriminability when either the inter-domain discrepancy is substantial or a severe distribution gap occurs between source and target domains.

Thirdly, the proposed UPDA framework substantially enhances cross-dataset generalization across all benchmark methods.
Notably, compared with NoAdapt, GMS-3DQA \cite{zhang2024gms} with UPDA achieves a large margin of 0.2241/0.1328 on average SRCC/PLCC during SJTU-PCQA to WPC 2.0 transfer.
The 3DTA \cite{zhu20243dta} with UPDA exhibits similar improvements during SJTU-PCQA to WPC 2.0 transfer, with a gain of 0.1225/0.1200 on average SRCC/PLCC compared with NoAdapt.

\begin{figure}[!tbp]
    \centering   %居中放置
    \subfloat[NoAdapt] %为每个图片加上编号
    {
        \begin{minipage}[b]{0.45\linewidth} 
            \centering
            \includegraphics[scale=0.2]{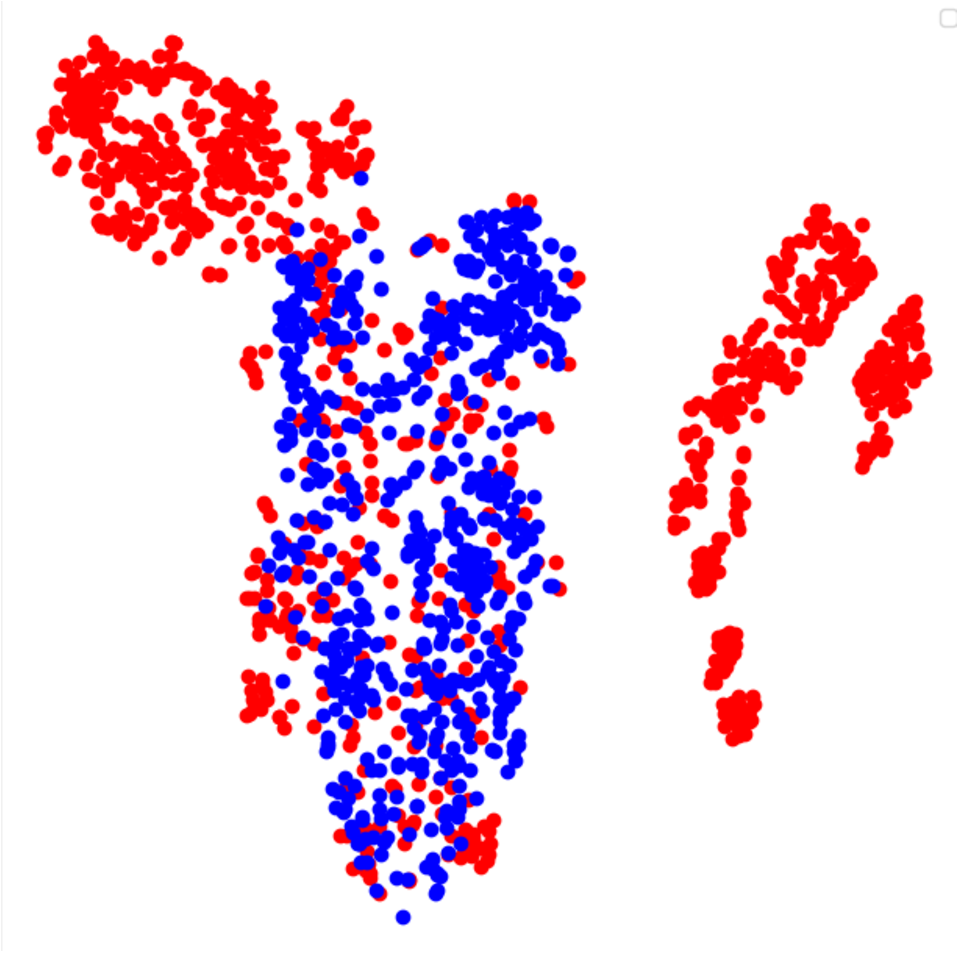}
        \end{minipage}
    }
    \subfloat[UPDA]
    {
        \begin{minipage}[b]{0.45\linewidth}
            \centering
            \includegraphics[scale=0.2]{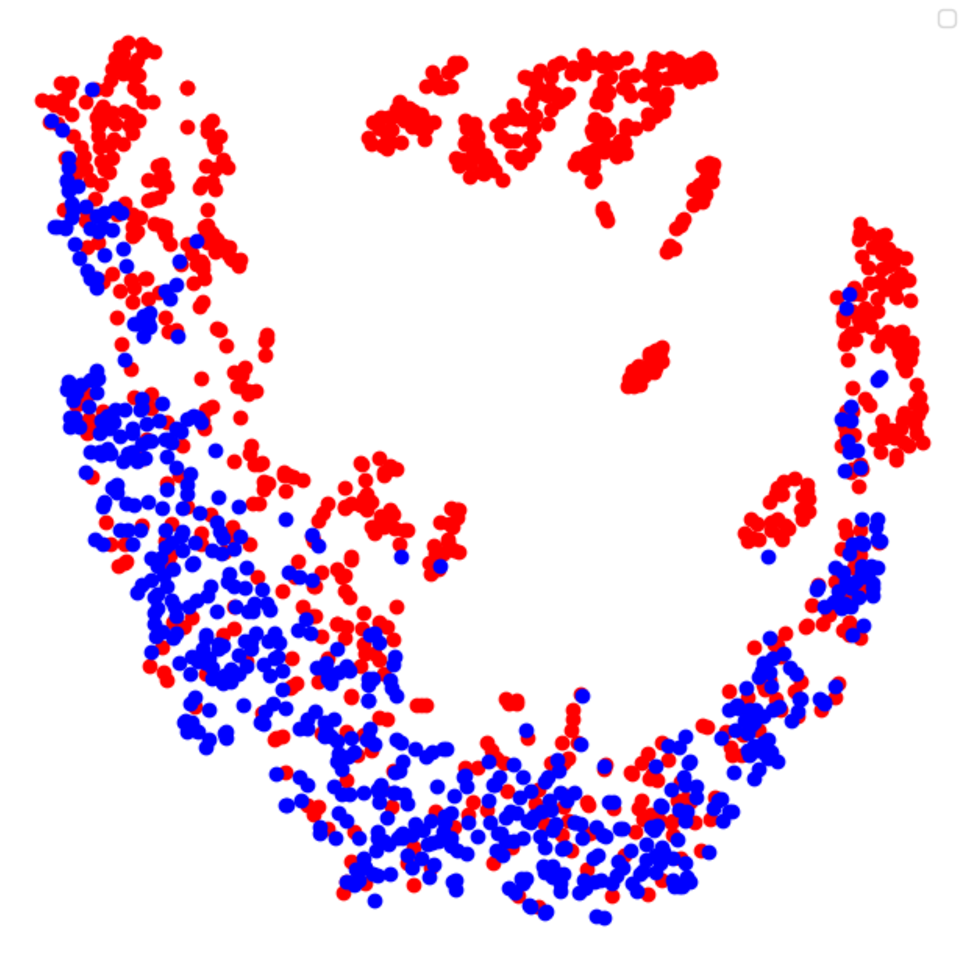}
        \end{minipage}
    }
    \caption{{t-SNE visualization of the feature space (red: source domain feature, blue: target domain feature).}}
    \label{fig:tsne}
\end{figure}

Furthermore, experimental results reveal only marginal performance gains when applying UPDA to the WPC and WPC 2.0 adaptation tasks. 
This limited improvement stems from the high intrinsic similarity between these datasets, particularly in: (1) shared distortion characteristics (notably V-PCC artifacts), and (2) nearly identical content distributions. 
Such minimal domain discrepancy reduces the opportunity for meaningful feature space alignment, consequently constraining the potential performance enhancement of UPDA in this specific configuration.
{We conduct t-SNE visualization experiments on the experimental setup of WPC/WPC 2.0 based 3DTA \cite{zhu20243dta}.
As shown in Fig. \ref{fig:tsne}, visual analysis reveals a significant overlap between the feature distributions of the source and target domains, resulting in an insignificant reduction in the domain difference after UPDA, which may weaken the UPDA's effectiveness.
}

\subsection{{Parameter Sensitivity Analysis}}

\begin{figure}[!tbp]
    \centering   %居中放置
    \subfloat[Parameter sensitivity analysis of $\nu$] %为每个图片加上编号
    {
        \begin{minipage}[b]{\linewidth} 
            \centering
            \includegraphics[scale=0.5]{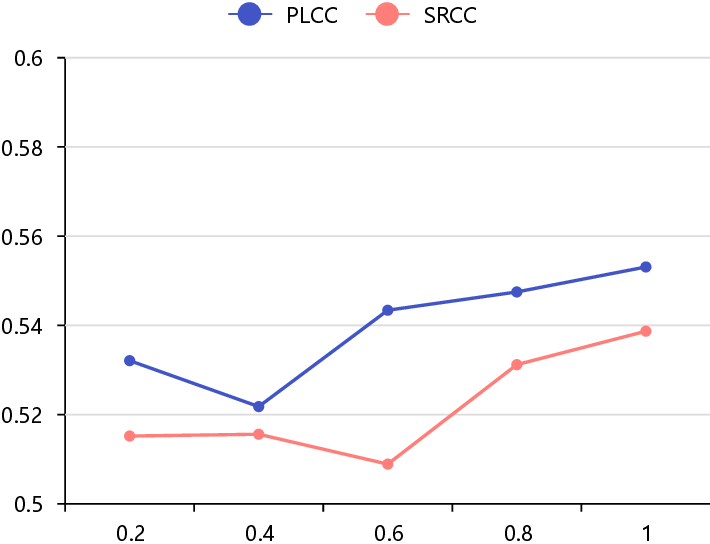}
        \end{minipage}
    }
    
    \subfloat[Parameter sensitivity analysis of $\mu$]
    {
        \begin{minipage}[b]{\linewidth}
            \centering
            \includegraphics[scale=0.5]{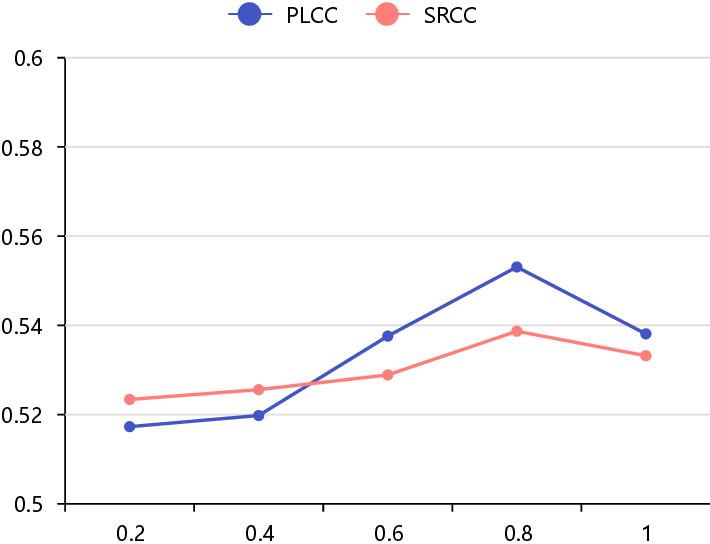}
        \end{minipage}
    }
    \caption{{Parameter sensitivity analysis of $\nu$ and $\mu$. The baseline model is GMS-3DQA \cite{zhang2024gms}.}}
    \label{fig:nu_mu}
\end{figure}

{To analyze the impact of weighting factors $\nu$ and $\mu$, we conduct the parameter sensitivity experiment.
Specifically, the values of $\nu$ and $\mu$ are taken within the range of $\{0.2, 0.4, 0.6, 0.8, 1\}$.
Multiple experiments are conducted to record the changes in the predictive indicators PLCC and SRCC of the model in the target domain. 
First, we use the regression head trained in the source domain to find a suitable $\nu$.
After determining the value of $\nu$, we conduct parameter sensitivity experiments on $\mu$.
It can be seen in Fig. \ref{fig:nu_mu} that the model achieves the best predictive performance in the target domain when $\nu=1$ and $\mu=0.8$, respectively.}

\subsection{Ablation Study}

In this section, we conduct a comprehensive ablation study to evaluate the individual contributions of UPDA's core components: (1) discrepancy-aware coarse-grained alignment (DACA), (2) Symmetric Feature Fusion (SFF), and (3) Conditional Discriminator (CD). 
To ensure rigorous evaluation under substantial domain shifts, we employ three challenging cross-dataset configurations (SJTU-PCQA/WPC, WPC/SJTU-PCQA, and WPC 2.0/SJTU-PCQA), deliberately excluding the more similar WPC/WPC 2.0 pairing. 
The quantitative results, presented in Table \ref{tab:ablation}, demonstrate the distinct impact of each component on model performance.

\subsubsection{The Effect of Discrepancy-aware Coarse-grained Alignment}

The experimental results indicate that removing DACA leads to a significant performance decrease of all backbones in cross-domain scenarios. 
In particular, when evaluated cross-domain from SJTU-PCQA to WPC, the MM-PCQA \cite{zhang2023mm} exhibits severe performance degradation: SRCC drops from 0.2381 to 0.1542 (a decrease of 35.2\%), and its PLCC collapses from 0.4659 to 0.1426 (a decrease of 69.4\%).
These results demonstrate that DACA successfully circumvents the challenges inherent in direct absolute feature alignment by instead learning transferable relative quality rankings across domains. 
Specifically, by modeling pairwise ranking relationships between cross-domain point clouds, DACA preserves quality-discriminative patterns while remaining robust to domain-specific feature variations.

% \usepackage{multirow}
\begin{table*}[tbp]
\centering
\caption{Ablation studies of discrepancy-aware coarse-grained alignment (DACA), symmetric feature fusion (SFF), and conditional discriminator (CD).}
\begin{tabular}{lccccccccccc}
\hline
 Train/Test &     &    &    & \multicolumn{2}{c}{SJTU-PCQA/WPC} & \multicolumn{2}{c}{WPC/SJTU-PCQA}  & \multicolumn{2}{c}{WPC 2.0/SJTU-PCQA} & \multicolumn{2}{c}{Average}\\
 Backbone & DACA & SFF & CD & SRCC        & PLCC       & SRCC        & PLCC       & SRCC         & PLCC     & SRCC         & PLCC   \\ \hline
\multirow{4}{*}{GMS-3DQA \cite{zhang2024gms}}  &  \XSolid   & \Checkmark  & \Checkmark  & 0.5367      & 0.5473     & 0.7699      & 0.8285     & 0.6752       & 0.7314  &   0.6606 &   0.7021  \\
&\Checkmark   & \XSolid  & \Checkmark  & 0.5273      & 0.5298     & 0.7478      & 0.7567     & 0.6352       & 0.7018  &  0.6368  & 0.6623    \\
&\Checkmark   & \Checkmark  & \XSolid  & 0.5301      & 0.5321     & 0.7566      & 0.7831     & 0.6683       & 0.6872  & 0.6516  & 0.6675    \\
&\Checkmark   & \Checkmark  & \Checkmark   & \textbf{0.5397}     & \textbf{0.5542 }    & \textbf{0.8167}        & \textbf{0.8354}         & \textbf{0.7398 }       & \textbf{0.7753 }   & \textbf{0.6987 }       & \textbf{0.7216 }       \\\hdashline
\multirow{4}{*}{3DTA \cite{zhu20243dta}}  & \XSolid   & \Checkmark  & \Checkmark     & 0.4319      & 0.5011     & 0.7025      & 0.7509     & 0.6778       & 0.6644  & 0.6041 & 0.6388    \\
& \Checkmark   & \XSolid  & \Checkmark      & 0.4476      & 0.5241     & 0.7653      & 0.7783     & 0.7012       & 0.6852   & 0.6380    & 0.6625   \\
& \Checkmark   & \Checkmark  & \XSolid       & 0.4542      & 0.5325     & \textbf{0.7851}      & 0.8212     & 0.7213       & 0.7277  & 0.6535    & 0.6871    \\
& \Checkmark   & \Checkmark  & \Checkmark    &\textbf{ 0.4814 }    & \textbf{0.5483}        & 0.7819        & \textbf{0.8189}     & \textbf{0.7523}        & \textbf{0.7401}        & \textbf{0.6720}     & \textbf{0.7024 }    \\ \hdashline
\multirow{4}{*}{MM-PCQA \cite{zhang2023mm}}  &  \XSolid   & \Checkmark  & \Checkmark   & 0.1542      & 0.1426     & 0.7892      & 0.8182     & 0.6764       & 0.6893      & 0.5399 & 0.5500    \\
&  \Checkmark   & \XSolid  & \Checkmark    & 0.1788      & 0.3019     & 0.7652      & 0.7931     & 0.6473       & 0.7212      & 0.5304  & 0.6054         \\
&  \Checkmark   & \Checkmark  & \XSolid   & 0.1882      & 0.4018     & 0.8019      & 0.8117     & 0.6908       & \textbf{0.7961}      & 0.5603 & 0.6698      \\
&  \Checkmark   & \Checkmark  & \Checkmark    &  \textbf{0.2381}     & \textbf{0.4659}       & \textbf{0.8173}        & \textbf{0.8283 }      & \textbf{0.6937}        & 0.7842        & \textbf{0.5830}     & \textbf{0.6928 }    \\  \hline
\end{tabular}
\label{tab:ablation}
\end{table*}

\subsubsection{The Effect of Symmetric Feature Fusion}

The removal of the symmetric feature fusion (SFF) module leads to a significant degradation of the cross-domain generalization ability of backbones. 
Taking the GMS-3DQA \cite{zhang2024gms} as an example, in the WPC/SJTU-PCQA test, after removing SFF, the SRCC decreases from 0.8167 to 0.7478 (a decrease of 8.4\%), and the PLCC decreases from 0.8354 to 0.7567 (a decrease of 9.4\%). 
These results demonstrate that the SFF effectively fuses quality-sensitive features across domains via a symmetric cross-attention mechanism, which dynamically captures shared structural distortions (e.g., geometric noise) and domain-specific artifacts (e.g., compression patterns).

\subsubsection{The Effect of Conditional Discriminator}

The absence of the conditional discriminator (CD) module significantly reduces the generalization ability of backbones.
Taking the 3DTA \cite{zhu20243dta} as an example, after removing CD, the average SRCC decreases by 2.7\% (from 0.6720 to 0.6535), and the PLCC decreases by 2.2\% (from 0.7024 to 0.6871).
The CD effectively reduces the effect of negative features on the feature alignment process through adversarial training and feature selection mechanisms. 
In the process of cross-domain adaptation, the model can focus more on features that are helpful for quality prediction, thereby improving the quality of feature alignment. 
This module enables the model to maintain a more stable performance output in the face of complex and diverse cross-domain data and enhances the reliability of the model in practical applications.
An abnormal phenomenon is that the MM-PCQA \cite{zhang2023mm} shows performance improvement after removing CD in the experimental setup of WPC 2.0/SJTU. 
One possible reason is that MM-PCQA \cite{zhang2023mm} involves multimodal feature fusion, and the fused features contain multiple types of information, which can make it difficult for CD to distinguish quality-related information.

\subsection{Computational Efficiency}

{
To analyze the computational efficiency of the proposed UPDA, this section conducts a comparative experiment on the computational overhead of its training phase, taking the GMS-3DQA \cite{zhang2024gms} as an example.
As shown in Table \ref{tab:efficiency}, the UPDA introduces a modest increase in computational cost during the training phase (about 10.5\% in FLOPs), which remains within acceptable limits for practical deployment.
In fact, as an offline adaptation method, the proposed UPDA does not introduce additional computational overhead in the inferencing phase.
}

\begin{table}[!tbp]
\caption{{Computational efficiency analysis of different adaptive methods.}}
\centering
\begin{tabular}{lccc}
\hline
Backbone                  & Method   & Params (M) & FLOPs (G)  \\ \hline
\multirow{3}{*}{GMS-3DQA} & NoAdapt  & 27.55 & 13.14 \\
                          & DirAdapt & 27.78 & 13.27 \\
                          & UPDA     & 34.58 & 14.52 \\
\hline
\end{tabular}
\label{tab:efficiency}
\end{table}

\section{Conclusion and Future Work} 
\label{sec:conclusion}

This paper presents an unsupervised progressive domain adaptation (UPDA) framework for cross-domain point cloud quality assessment, comprising two key components: discrepancy-aware coarse-grained alignment and perception fusion fine-grained alignment. 
While unsupervised domain adaptation has been extensively studied for image quality assessment, to the best of our knowledge, UPDA represents the first work to systematically address domain shifts in PCQA through a hierarchical coarse-to-fine alignment strategy. 
Notably, UPDA achieves superior cross-domain generalization without modifying the base network architecture, making it particularly suitable for real-world applications. 
{Although this work focuses on unsupervised domain adaptation for point cloud quality assessment, future research directions can explore source-free domain adaptation and test time adaptation to face more realistic scenarios.}
% While this work focuses on unsupervised domain adaptation for point cloud quality assessment, future research directions could explore source-free domain adaptation (SFDA) scenarios where source domain data becomes inaccessible during adaptation—a practical constraint often encountered in real-world deployments.

% \section*{Acknowledgments}
% This work was supported in part by National Key R\&D Program of China under Grant  2022QY0102, and Pre-research Project of The 14th Five Year Plan under Grant 50904040201 and Grant 2020-JCJQ-ZD-007-00.

\bibliographystyle{IEEEtran}
\bibliography{main}

% \end{IEEEbiography}

\begin{IEEEbiography}[{\includegraphics[width=1in,height=1.25in,clip,keepaspectratio]{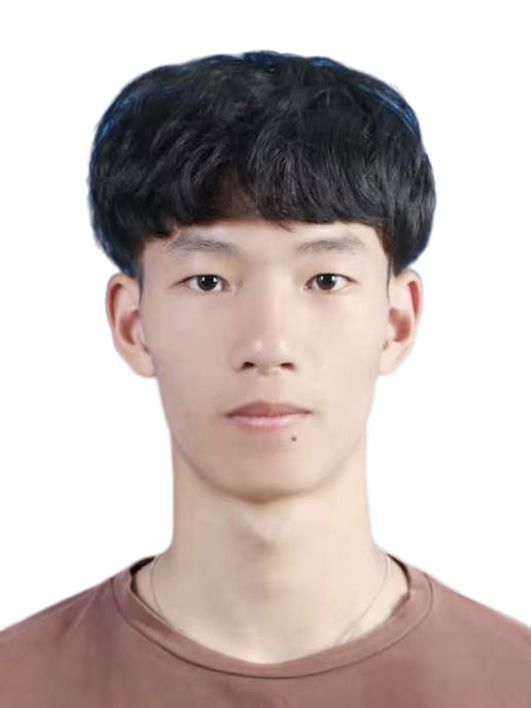}}]{Bingxu Xie} received the B.S. degree from Jiangnan University, Jiangsu, China in 2023 and now studies at Nanjing University of Science and Technology. His research interests include point cloud quality assessment and machine learning.
\end{IEEEbiography}

\begin{IEEEbiography}[{\includegraphics[width=1in,height=1.25in,clip,keepaspectratio]{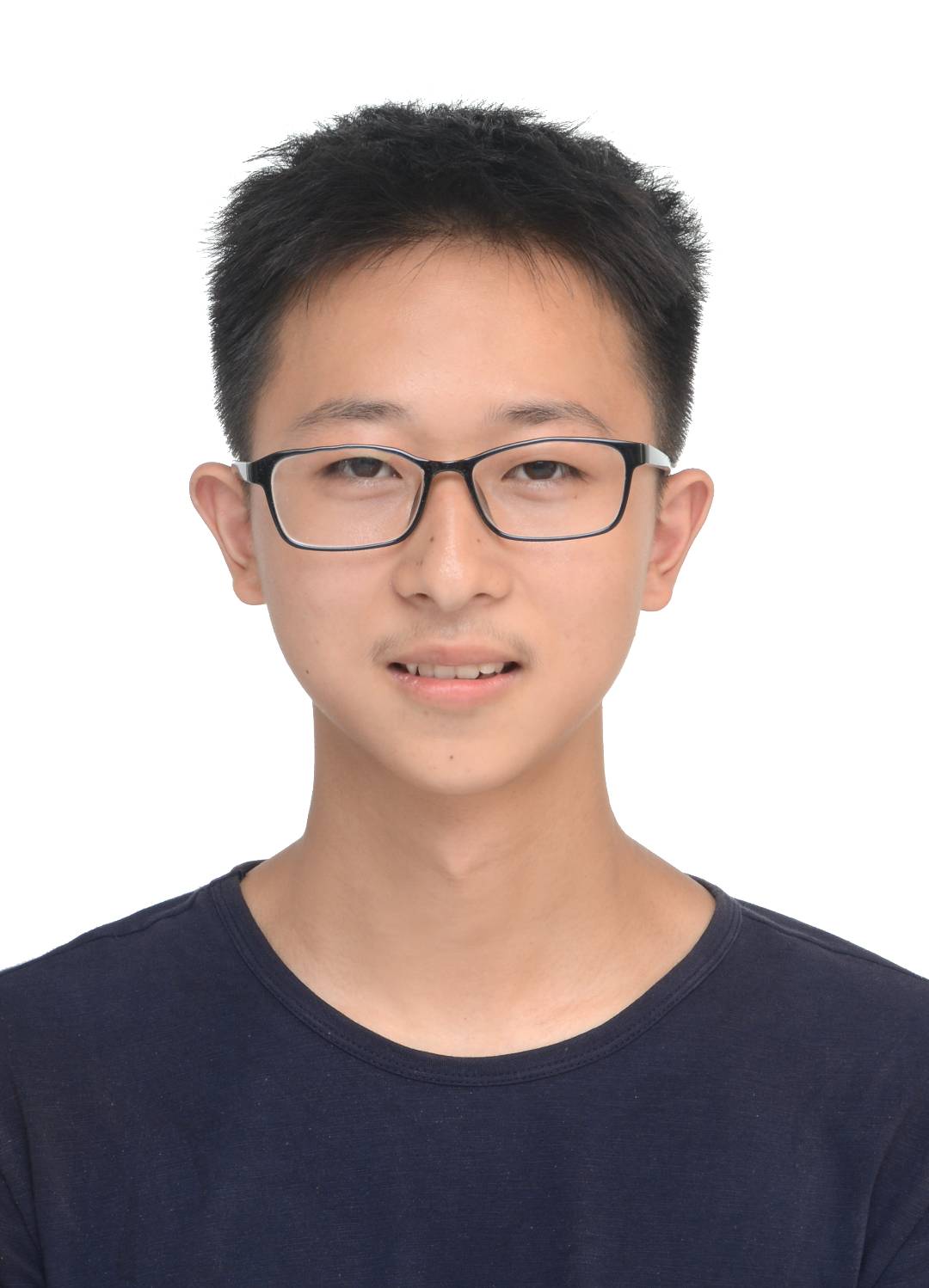}}]{Jincan Wu}
received the B.S. degree from Nanjing University of Science and Technology, Jiangsu, China in 2023 and now studies at Nanjing University of Science and Technology. His research interests include mesh quality assessment and 3DGS quality assessment.
\end{IEEEbiography}

\begin{IEEEbiography}[{\includegraphics[width=1in,height=1.25in,clip,keepaspectratio]{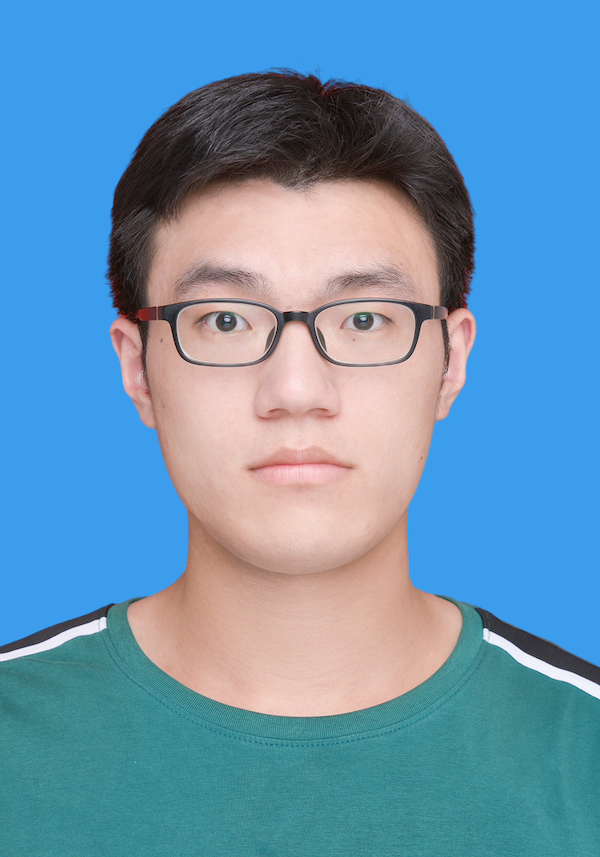}}]{Yonghui Liu}
received the B.S. degree from China Jiliang University, Hangzhou, China in 2024 and now studies at Nanjing University of Science and Technology. His research interests include infrared image simulation and image quality assessment.
\end{IEEEbiography}

\begin{IEEEbiography}[{\includegraphics[width=1in,height=1.25in,clip,keepaspectratio]
{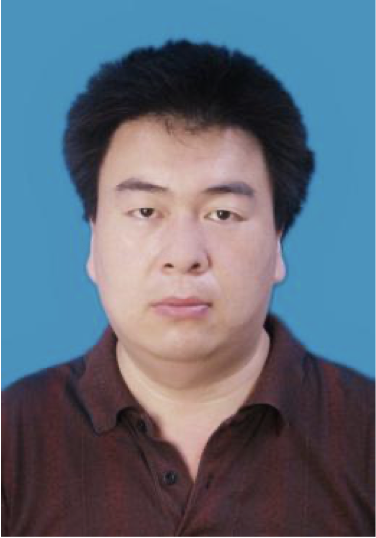}}]{Weiqing Li}
is currently an associate professor at the School of Computer Science and Engineering, Nanjing University of Science and Technology, China. He received the B.S. and Ph.D. degrees from the School of Computer Sciences and Engineering, Nanjing University of Science and Technology in 1997 and 2007, respectively. His current interests include computer graphics and virtual reality.
\end{IEEEbiography}

\begin{IEEEbiography}[{\includegraphics[width=1in,height=1.25in,clip,keepaspectratio]
{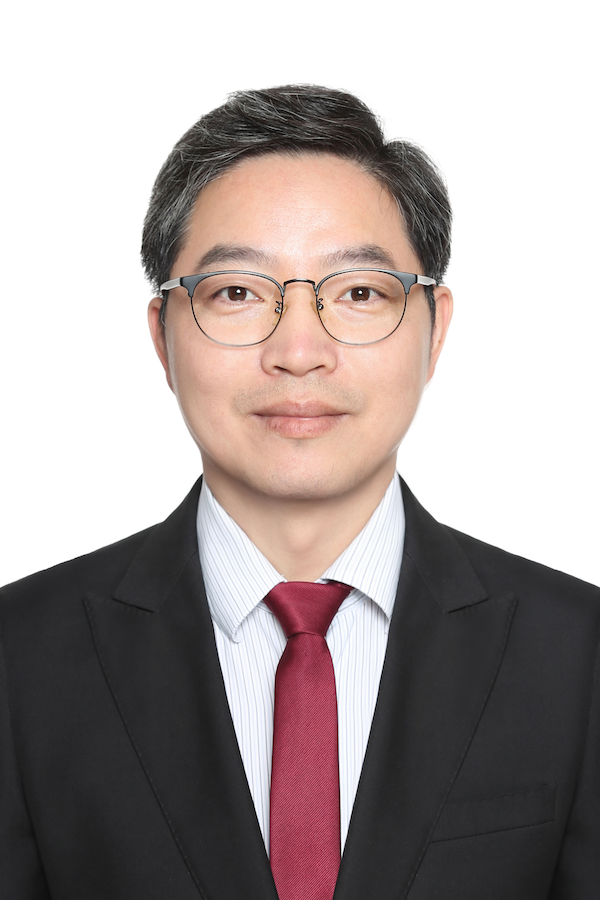}}]{Zhiyong Su}
is currently an associate professor at the School of Automation, Nanjing University of Science and Technology, China. He received the B.S. and M.S. degrees from the School of Computer Science and Technology, Nanjing University of Science and Technology in 2004 and 2006, respectively, and received the Ph.D. from the Institute of Computing Technology, Chinese Academy of Sciences in 2009. His current interests include computer graphics, computer vision, augmented reality, and machine learning.
\end{IEEEbiography}

\end{document}